\documentclass[sigconf]{acmart}

\AtBeginDocument{%
  \providecommand\BibTeX{{%
    \normalfont B\kern-0.5em{\scshape i\kern-0.25em b}\kern-0.8em\TeX}}}

\copyrightyear{2023}
\acmYear{2023}
\setcopyright{rightsretained}
\acmConference[RAID '23]{The 26th International Symposium on Research in Attacks, Intrusions and Defenses}{October 16--18, 2023}{Hong Kong, Hong Kong}
\acmBooktitle{The 26th International Symposium on Research in Attacks, Intrusions and Defenses (RAID '23), October 16--18, 2023, Hong Kong, Hong Kong}
\acmDOI{10.1145/3607199.3607243}
\acmISBN{979-8-4007-0765-0/23/10}

\usepackage{algorithmic}
\usepackage{textcomp}
\usepackage{xcolor}
\usepackage[acronym,shortcuts]{glossaries}
\usepackage{soul}
\usepackage{float}
\usepackage{subcaption}
\usepackage{url}
\usepackage{stfloats}

\newacronym{psd}{PSD}{Personal Security Device}
\newacronym{rfid}{RFID}{Radio-Frequency IDentification}
\newacronym{rf}{RF}{Radio-Frequency}
\newacronym{em}{EM}{electromagnetic}
\newacronym{ask}{ASK}{Amplitude Shift Keying}
\newacronym{wgn}{WGN}{White Gaussian Noise}
\newacronym{pll}{PLL}{Phase-Locked Loop}
\newacronym{uid}{UID}{Unique Identifier}
\newacronym{pdf}{PDF}{Probability Density Function}
\newacronym{nfc}{NFC}{Near Field Communication}
\newacronym{std}{STD}{Standard Deviation}
\newacronym{prng}{PRNG}{Psuedo-Random Number Generator}
\newacronym{lfsr}{LFSR}{Linear Feedback Shift Register}
\newacronym{mfcuk}{MFCUK}{MIFARE Classic Universal toolKit}
\newacronym{fft}{FFT}{Fast Fourier Transform}
\newacronym{asr}{ASR}{Attack Success Rate}
\newacronym{ic}{IC}{Integrated Circuit}

\newcommand{\parag}[1]{\noindent\textbf{#1. }}

\begin{document}
\title[Beware of Pickpockets: A Practical Attack against Blocking Cards]{Beware of Pickpockets: \\A Practical Attack against Blocking Cards}

\author{Marco Alecci}
\orcid{0000-0002-5963-4599}
\affiliation{%
  \institution{SnT, University of Luxembourg}
  \city{Luxembourg}
  \country{Luxembourg}
}
\email{marco.alecci@uni.lu}

\author{Luca Attanasio}
\orcid{0000-0003-3653-7080}
\affiliation{%
  \institution{Department of Mathematics, University of Padova}
  \city{Padua}
  \country{Italy}
}
\email{luca_attanasio@me.com}

\author{Alessandro Brighente}
\orcid{0000-0001-6138-2995}
\affiliation{%
  \institution{Department of Mathematics, University of Padova}
  \city{Padua}
  \country{Italy}
}
\email{alessandro.brighente@unipd.it}

\author{Mauro Conti}
\orcid{0000-0002-3612-1934}
\affiliation{%
  \institution{Department of Mathematics, University of Padova}
  \city{Padua}
  \country{Italy}
}
\email{conti@math.unipd.it}
 
 \author{Eleonora Losiouk}
\orcid{0000-0002-2315-7823}
\affiliation{%
  \institution{Department of Mathematics, University of Padova}
  \city{Padua}
  \country{Italy}
}
\email{elosiouk@math.unipd.it}
 
 \author{Hideki Ochiai}
 \orcid{0000-0001-9303-5250}
\affiliation{%
 \institution{Department of Electrical and Computer Engineering, Yokohama National University}
 \streetaddress{Rono-Hills}
 \city{Yokohama}
 \country{Japan}}
\email{hideki@ynu.ac.jp}
 
\author{Federico Turrin}
 \orcid{0000-0001-5660-2447}
\affiliation{%
  \institution{Department of Mathematics, University of Padova}
  \city{Padua}
  \country{Italy}
}
\email{turrin@math.unipd.it}

\renewcommand{\shortauthors}{Alecci and Attanansio, et al.}

\begin{abstract}
Today, we rely on contactless smart cards to perform several critical operations (e.g., payments and accessing buildings). Attacking smart cards can have severe consequences, such as losing money or leaking sensitive information. Although the security protections embedded in smart cards have evolved over the years, those with weak security properties are still commonly used. Among the different solutions, blocking cards are affordable devices to protect smart cards. These devices are placed close to the smart cards, generating a noisy jamming signal or shielding them. Whereas vendors claim the reliability of their blocking cards, no previous study has ever focused on evaluating their effectiveness. 

In this paper, we shed light on the security threats on smart cards in the presence of blocking cards, showing the possibility of being bypassed by an attacker. We analyze blocking cards by inspecting their emitted signal and assessing a vulnerability in their internal design. We propose a novel attack that bypasses the jamming signal emitted by a blocking card and reads the content of the smart card. 

We evaluate the effectiveness of $11$ blocking cards when protecting a MIFARE Ultralight smart card and a MIFARE Classic card. Of these $11$ cards, we managed to bypass $8$ of them and successfully dump the content of a smart card despite the presence of the blocking card.
Our findings highlight that the noise type implemented by the blocking cards highly affects the protection level achieved by them. Based on this observation, we propose a countermeasure that may lead to the design of effective blocking cards. To further improve security, we released the tool we developed to inspect the spectrum emitted by blocking cards and set up our attack.
\end{abstract}



\maketitle

\section{Introduction}\label{sec:intro}
Today, \emph{smart cards} represent an enabling technology to perform several critical operations like cashless payments, access control, employee IDs, and e-passports. Smart cards are physical cards that embed an~\ac{ic} chip to store and process data. They can communicate with a reader through physical contact or short-range wireless protocols. In particular, contactless smart cards rely on the \ac{rfid} standard to interact with a reader without requiring the card to be physically inserted. The use of contactless smart cards has been growing in recent years: in 2021, more than 80\% of US consumers relied on contactless smart cards, while between 2019 and 2020, there has been a 150\% increase in contactless payment transactions~\cite{contactless_stats}. Attackers have widely targeted smart cards~\cite{gupta2021taxonomy} through active or passive attacks. The former involves physical access to hardware, leading to probing~\cite{Shi2017} and reverse engineering~\cite{courbon2016reverse}, while the latter focuses on information leakages, such as power~\cite{mangard2008power} and timing~\cite{kocher1996timing} side-channels. Finally, other attacks are executed over the air by relaying~\cite{hancke2005practical}, eavesdropping~\cite{kortvedt2009eavesdropping}, and skimming~\cite{hancke2011practical}. Attacks against smart cards have a huge impact: according to the Federal Trade Commission, credit card threats, like identity theft, fraud, and data breaches, have affected 390 million people in the United States in 2021, while identity theft doubled between 2019 and 2020 and has increased in 2021, affecting approximately 1.7 million people~\cite{fraudstats}. Both identity cards and credit cards are high-level implementations that, under the hood, use the same protocols discussed in this paper.
The level of cryptographic security in both smartcards and the associated protocols is significantly inadequate. It is weak to such an extent that an attacker communicating with a smartcard like a regular payment terminal or door lock can typically extract enough data to produce a counterfeit replica of the card~\cite{5207633}.

Two alternative approaches can be identified to defend against the attacks mentioned above: enhancing the inner security of smart cards; introducing an external component to protect them. Considering the former, the ISO/IEC 14443~\cite{ISO/IEC14443-1} standard describes the physical characteristics of smart cards available on the market, of which MIFARE is one of the biggest players. According to the specification, there are four types of MIFARE cards, each with an increased security level: (i) the MIFARE Ultralight comes with a $32$-bit password, (ii) the MIFARE Classic relies on authentication and encryption methods, being its memory sectors protected by $48$-bit keys, (iii) MIFARE Plus uses a $128$-bit AES encryption scheme, and (iv) MIFARE DESFire embeds cryptographic algorithms such as symmetric Triple-DES or AES. Although the last two types are used for systems requiring more robust security (e.g., electronic payments, e-passports, identity cards), MIFARE Ultralight and MIFARE Classic are widely used in access control systems, such as public transportation, event ticketing, prepaid applications, loyalty, and amusement~\cite{mifareapplication}. In these scenarios, users can rely on the security introduced by an external component to protect their smart cards, such as blocking cards, blocking wallets, or blocking covers. While covers and wallets generally rely on their metal shielding structure to protect smart cards, blocking cards employ a higher range of approaches (e.g., shielding, jamming), making them more interesting and challenging to study from a research point of view.

Blocking cards offer an affordable solution to prevent attacks performed over wireless communications against smart cards. Placed near smart cards, blocking cards protect them through a passive or active approach. On the one hand, passive protection is provided by shielding the communication or emitting a jamming signal based on a received stimulus of the specific carrier frequency. On the other hand, active protection involves continuous noise generated without a stimulus.
Even though blocking cards are one of the possible countermeasures users can implement, vendors can provide very little information as they wish to keep their internal design secret. 

In this work, we perform the first security evaluation of blocking cards. We first analyze their emitted spectrum and identify that most blocking cards emit uncorrelated jamming signals with a Gaussian mixture distribution. 
We hence demonstrate the vulnerability we found by designing an attack aimed at exfiltrating the content of a smart card when protected by such blocking cards. We assume that the attacker interacts with the victim's smart card and intercepts the compound signal generated by the blocking card and the smart card by being physically close to the victim. The attacker then elaborates on the received signal to remove the noise and extract the content of the smart card. We apply our attack in a proof-of-concept scenario, evaluating the effectiveness of $11$ blocking cards in protecting MIFARE Ultralight and MIFARE Classic smart cards. 
By analyzing the emitted spectrum, we have determined that the $11$ blocking cards are reactive, and we have effectively executed our attack against $8$ of them. After identifying the limitations of the current blocking cards' internal design, we experimentally evaluate the performance of different types of noise to find the ones that effectively protect smart cards. We believe this study will help vendors strengthen the robustness of their developing blocking cards. Finally, we develop and release a tool that carries out our attack and implements our countermeasure. 

. 

Our contributions are as follows:
\begin{itemize}
    \item We perform a security analysis of $11$ popular blocking cards at the time of conducting our experiment;
    \item We design a novel attack aimed at extracting the content of a smart card that is protected by a blocking card. We execute the attack against $11$ blocking cards protecting MIFARE cards and evaluate their vulnerability. Our results highlight that the noise type implemented by the blocking cards highly affects the eavesdropping protection quality;
    \item We provide an in-depth analysis of the effectiveness of different types of noise, which may serve as a design guideline for the future development of blocking cards;
    \item We release our tool in open source at:
    \begin{center}
    \url{https://github.com/spritz-group/BlockingCardAnalysis}
    \end{center}
\end{itemize}

\parag{Responsible Disclosure}
At the end of our security analysis, in October 2022, we informed the vendors of $6$ blocking cards ($\approx43\%$) about our findings. We could not find contact information for $5$ of them ($36\%$).
We received a response from $4$ out of the $6$ we contacted: $2$ of them asked not to disclose their brand, and the other $2$ did not respond after we shared the vulnerability details. Thus,  we decided to anonymize all the blocking card brands to have a uniform approach toward them all and to prevent any brand disclosure issues.

We believe that the reasons for such a low response rate may be two-fold. First, the low price of most blocking cards (i.e., less than \textdollar$10$ for $10$ of them) makes them competitive on the market but discourages their vendors from introducing extra security features. Second, blocking cards come with limited hardware, which might make implementing stronger approaches (e.g., jamming signals or ad-hoc jamming patterns) difficult.

\section{Background on Smart cards}\label{sec:smart_cards}
In this section, we first provide an overview of smart card technology (i.e., Section~\ref{sec:general}), then we introduce the two smart cards we considered in our scenario: the MIFARE Ultralight (i.e., Section~\ref{subsec:back_ultralight}) and the MIFARE Classic (i.e., Section~\ref{subsec:back_classic}).

\subsection{Smart Cards Technology}\label{sec:general}
The standard ISO 7816-1~\cite{ISO7816-1:2011} describe smart card with contact technology. This class of cards includes different technologies: magnetic stripe cards, contact smart cards, and proximity cards. While contact smart cards must be inserted into the reader to communicate with it, proximity cards rely on the energy transferred by the reader over~\ac{rfid} to power their microprocessor. In fact, proximity cards do not have an internal supply battery and are powered by the reader through \ac{em} field. This paper focuses on proximity cards; whenever we use the ``smart card'' terminology, we specifically refer to proximity cards. 

According to the ISO/IEC 14443~\cite{ISO/IEC14443-2} standard, the transmission carrier frequency between the card and the reader is $f_c=13.56$~MHz, and the reader magnetic field strength is at least $H_{min}=1.4 $~ A/m and at most, $H_{max} = 7.5$~A/m. The field strength of the card is approximately $1.5$~W/m, resulting in a maximum communication range of $10$~cm. The communication between reader and card proceeds as follows: the reader activates the card by applying an~\ac{em} field; the card waits for a command sent by the reader and when received, might transmit a response; the reader and the card start the communication; the reader deactivates the~\ac{rf} operating field. Since a reader might already communicate with a card, it relies on the ``anti-collision protocol'' to select which should receive the message.

\subsection{MIFARE Ultralight}\label{subsec:back_ultralight}

The MIFARE Ultralight is the simplest card belonging to the MIFARE family. The first MIFARE Ultralight version had 512 bits (16 pages of 4 bytes) of memory and no security protections.
In this paper, we focus on the MIFARE Ultralight EV1 model~\cite{mifareultralightev12014n}. This card has 1024-bit memory, prevents the rewriting of memory pages through One-Time-Programmable bits and a write-lock, and guarantees data access protection through a 32-bit password. It is generally used in application scenarios that do not involve sensitive data (for example, the cash balance~\cite{mifareapplication}).

\subsection{MIFARE Classic}\label{subsec:back_classic}
The MIFARE Classic card has more security features than the MIFARE Ultralight. It is based on an NXP Semiconductor proprietary security protocol called CRYPTO-1 for both authentication and encryption of data exchange~\cite{mifareclassic}. Among the different MIFARE Classic cards, we focus on EV1, which is the best in this family. The MIFARE Classic EV1 is available in 1K and 4K memory versions. In both versions, the memory is organized into sectors of four or more blocks of 16 bytes each.

In 2008, researchers reverse-engineered the MIFARE Classic chip and recovered the CRYPTO1 algorithm by slicing the chip and taking pictures with a microscope~\cite{Nohl2008ReverseEngineeringAC}.
In the same period, other researchers followed a software-oriented approach and recovered the logical description of the cipher and communication protocol~\cite{koning2008practical, garcia2008dismantling}. 
In particular, in~\cite{koning2008practical} Gans et al. studied the malleability of the CRYPTO1 stream cipher to read all memory blocks of the first sector of the card, while in~\cite{garcia2008dismantling} Garcia et al. reverse-engineered MIFARE Classic based on the communication behavior between a card and a reader. The result of these previous works is a complete reversal of both the authentication protocol and the encryption algorithm, which led to the identification of several vulnerabilities. The main one is the poor design of the \ac{prng} used by the card to generate the nonce to be sent to the reader since it is possible to predict the next nonce used by the card~\cite{garcia2008dismantling}.

\section{Related Work}\label{sec:related}

Since introducing Proximity cards for contactless payments in 2007, governments, companies, and researchers have been actively investigating transaction security~\cite{gupta2021taxonomy}.

\parag{Attacks on the Smart Card Hardware} The most common attacks that are effective against smart cards focus on hardware vulnerabilities. These attacks are carried out with techniques including probing~\cite{Shi2017}, reverse engineering~\cite{courbon2016reverse}, or power~\cite{mangard2008power} and timing~\cite{kocher1996timing} side-channel analysis. However, these attacks require a skilled attacker and a complicated attacker's model (e.g., expensive instrumentation and stealing a smart card).
Unlike the attacks mentioned above, the attack we propose focuses on the communication between the reader and the card.

\parag{Smart Card Communication Attacks} The most well-known communication protocol attack is the relay, first practically introduced by Hancke et al.~\cite{hancke2005practical}. This attack aims to transfer the entire communication flow from one payment terminal to another to fraud and charge the victim for a transaction.
In~\cite{hancke2011practical}, the authors show how performing a skimming attack on ~\ac{rfid} tokens is possible. Another common attack to steal private information stored on the smart card is the snooping attack~\cite{konidala2007simple}, which consists in accessing unauthorized data from another person or company (e.g., casual observance of the card's PIN). Other works analyze the authentication protocol of some smart cards, exposing their weaknesses.
In~\cite{garcia2008dismantling}, the authors successfully attack the authentication protocol of a MIFARE Classic card, while in~\cite{5207633}, the authors extend this attack by requiring only wireless access to the card without requiring any~\ac{rfid} reader. In~\cite{boureanu2018another}, the authors show the feasibility of relay attacks against the EMV payment protocol on smart cards.
As we did in this paper, none of the previous works uses a blocking card to protect the smart card.

\parag{Jamming} The jamming technique relies on generating radio frequencies to corrupt a wireless communication, either by keeping the medium busy or manipulating the signal received by the receivers. Jammers can rely on different approaches~\cite{jamming_survey}: proactive if the jamming signal is transmitted when data are in the network; reactive if the jamming signal is generated only when there are data in the network; function-specific if the jammer has a specific purpose. 
Although jamming is usually associated with malicious usage, such a technique also has possible benign applications. Among those, reactive blocking cards that emit a jamming signal are designed to disrupt the data transmitted by a smart card and prevent an attacker from reading them. Another usage of the jamming technique, which however, comes with limitations~\cite{anti_jamming2}, is the ``friendly jamming'' or ``co-operative jamming''~\cite{friendly_jamming1, friendly_jamming2}, where there is an agreement among the emitter, the receiver, and the jammer so that the jammed signal is still recognizable for the two components involved in the communication, but not for an external eavesdropper. Finally, several anti-jamming techniques~\cite{jamming_survey} have been proposed (e.g., spectrum spreading, frequency hopping). However, previous works~\cite{jamming_survey} stated the ineffectiveness of anti-jamming techniques against RFID systems, which also holds validity for blocking cards we consider in this paper. Thus, our study on the effectiveness of blocking cards is novel and will contribute to enhancing community knowledge.

\parag{NFC Communication Defence} Several researchers proposed mechanisms to secure the \ac{nfc} communication. These defenses are generally designed for smartphone devices because they leverage jamming.
In~\cite{gummeson2013engarde}, the authors propose \textit{EnGarde}, a hardware-level solution able to protect the smartphone from malicious \ac{nfc} communication. Similarly, in ~\cite{zhou2014nshield}, the authors illustrate a non-invasive hardware solution to protect the smartphone from eavesdropping attacks on \ac{nfc} communication. More recently, Di Pietro et al.~\cite{di2018n} propose a software-level solution to block unwanted \ac{nfc} communication with the smartphone. Although effective, these solutions are specifically designed for smartphones, restricting the portability of the approach. Instead, the jamming solution we study aims to improve the methods already used in blocking cards but can also be applied to similar hardware constraint devices.

\parag{Blocking Card Analysis} In~\cite{yt_2}, the video's presenter evaluated the effectiveness of a blocking card at different distances and positions between the reader and the smart card. The presenter also tried to disassemble it, although he was unsuccessful in his attempt. Our study involves a much more thorough approach since we receive and analyze the spectrum of the signal generated by blocking cards based on signal processing techniques. Furthermore, we successfully dissect a shielding card to examine its internal structure.

\section{Blocking Cards}\label{sec:blocking_card}
This section presents a possible taxonomy to classify blocking cards (i.e., Section~\ref{subsec:taxonomy}). We then illustrate how we classify our $14$ blocking cards by analyzing their emitted power spectrum (i.e., Section~\ref{subsec:class}) and their internal physical components (i.e., Section~\ref{subsec:shielding}).  

\subsection{Blocking Card Taxonomy}
\label{subsec:taxonomy}
Blocking cards provide the most affordable and common defense mechanisms to protect smart cards from over-the-air attacks. However, very little is known about them because their vendors keep their internal design secret, and the academic community lacks focus on them. In fact, there is no standard taxonomy for blocking cards in the literature, while we found some vendors and patents that refer to the same types of blocking cards. The first classification criterion for a blocking card is whether it is \emph{passive} or \emph{active}. In particular, passive ones retrieve the energy required for their activation from an external source, whereas active blocking cards come equipped with a battery that allows them to actively emit a jamming signal, which is usually stronger than the passive ones. Due to this choice of internal design, passive blocking cards are cheaper and more diffused than active ones. Furthermore, active blocking cards may even fail to comply with the regulations adopted by some countries due to the unauthorized signal transmission~\cite{powerreg}. Passive blocking cards can be distinguished further between \emph{shielding} or \emph{reactive}~\cite{audebert2006blocking} cards. The former are made of non-conducting materials, such as aluminum foils~\cite{shkolnikov2011shield}, to block the~\ac{em} field around the smart card by exploiting the Faraday cage principle (reactive and active cards, instead, disrupt the communication between the smart card and the ~\ac{rf} reader). Reactive blocking cards react after a stimulus is received from an \ac{nfc} reader at a given carrier frequency. Since they are not battery-powered, reactive blocking cards rely on the received \ac{rfid} energy to power up the jamming signal.

\subsection{Classification Based on Spectrum Analysis}
\label{subsec:class}
Following its internal recommendation system, we selected and bought the top 14 blocking cards from the Amazon marketplace. Each vendor claims on its website that the card implements RFID protection, particularly NFC communication protection (i.e., the operating frequency at $13.56$MHz, or HF RFID). Considering our $14$ blocking cards, we have no information on their internal design from the vendors, except for one of them being classified as active. Hence, we position each blocking card at about $4$cm from an~\ac{nfc} reader and record the emitted signal through a power spectrum analyzer (this distance is empirically identified to prevent saturation issues on the spectrum analyzer side). The recording is carried out with and without the~\ac{nfc} reader activated. In this way, we can discriminate between a continuously emitting active blocking card and a passive one. The analysis of the power spectrum of the signal generated by the blocking card allows inferring its internal design: reactive blocking cards generate noise only when there is an \ac{em} field; active blocking cards continuously generate noise, even without a \ac{rf} field, being battery powered; shielding blocking cards shield the communication, relying on the Faraday principle. Considering our $14$ blocking cards, we find that: 
\begin{itemize}
    \item $10$ are reactive. This is confirmed by the power level of the spectrum, which is low in the absence of a smart card and high in the presence of a smart card.
    \item $1$ is reactive, despite being sold as active.
    \item $3$ implement a shielding strategy.
\end{itemize}

\begin{figure*}[t]
\centering

\begin{subfigure}{0.47\textwidth}
\centering
\includegraphics[width=\columnwidth]{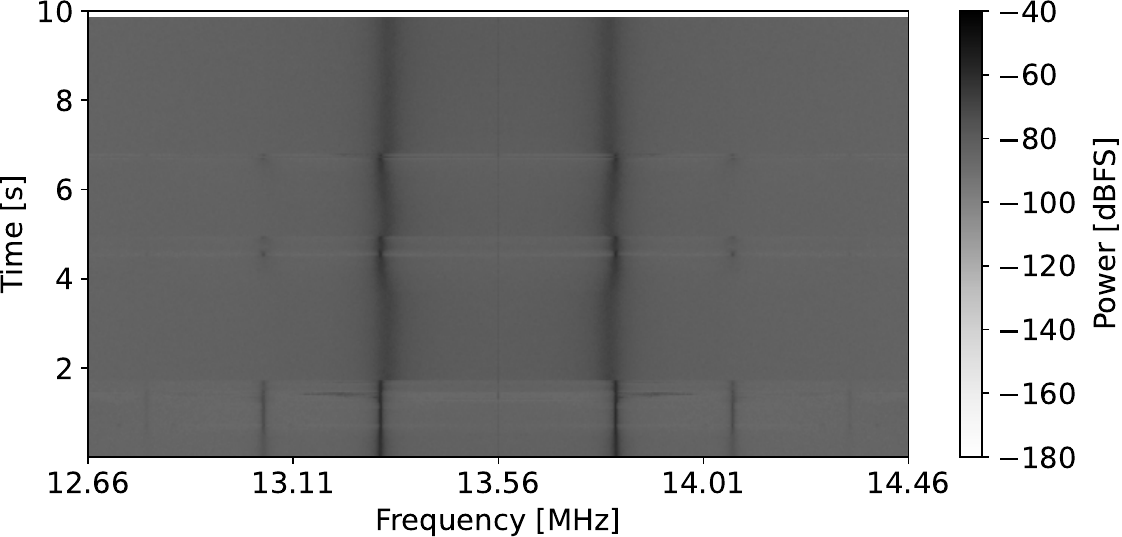}
\caption{Blocking Card 3.}\label{subfig:spectrum_bk3}
\end{subfigure}
\hfill
\begin{subfigure}{0.47\textwidth}
\centering
\includegraphics[width=\columnwidth]{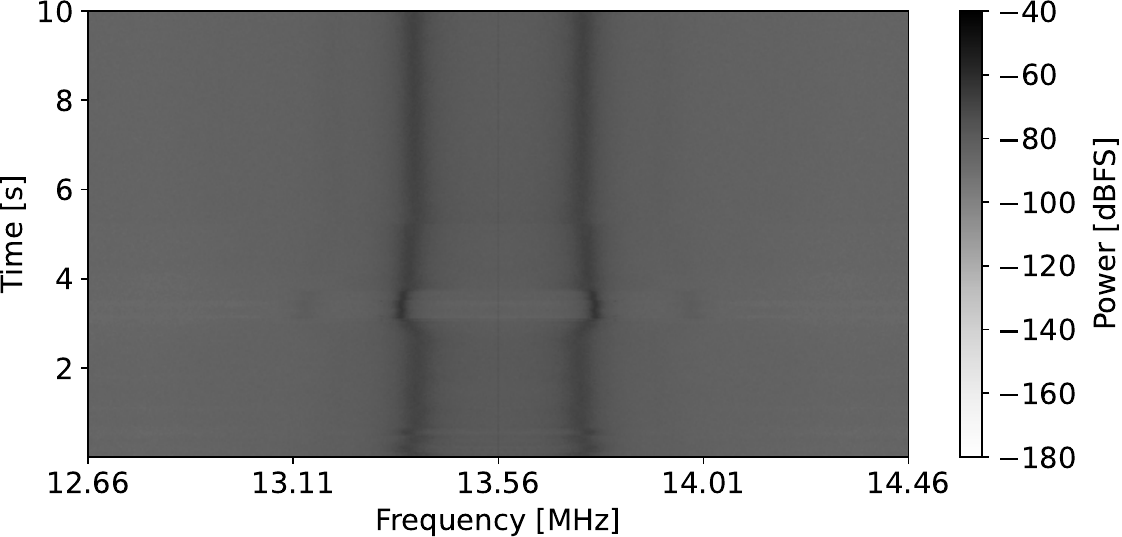}
\caption{Blocking Card 5.}\label{subfig:spectrum_bk5}
\end{subfigure}

\begin{subfigure}{0.47\textwidth}
\centering
\includegraphics[width=\columnwidth]{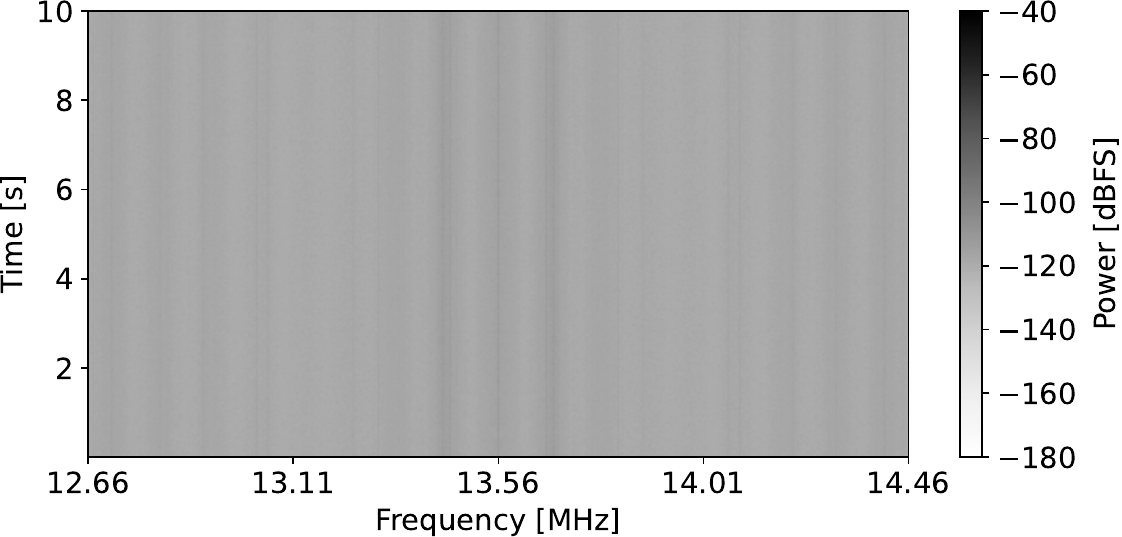}
\caption{Blocking Card 9.}\label{subfig:spectrum_bk9}
\end{subfigure}
\hfill
\begin{subfigure}{0.47\textwidth}
\centering
\includegraphics[width=\columnwidth]{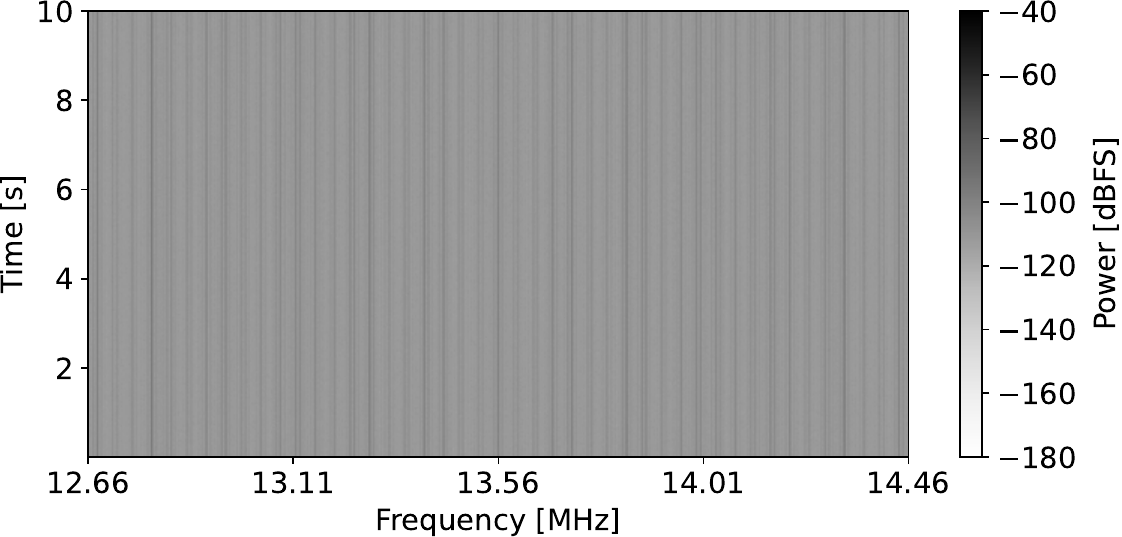}
\caption{Blocking Card 10.}\label{subfig:spectrum_bk10}
\end{subfigure}

\caption{Power spectrum of blocking cards emitting (a),(b) white noise and (c),(d) noise at multiple frequencies.}
\label{fig:spectrum}
\end{figure*}

\begin{figure}[t]
     \centering
     \begin{subfigure}{0.45\columnwidth}
     \centering\includegraphics[width=\columnwidth]{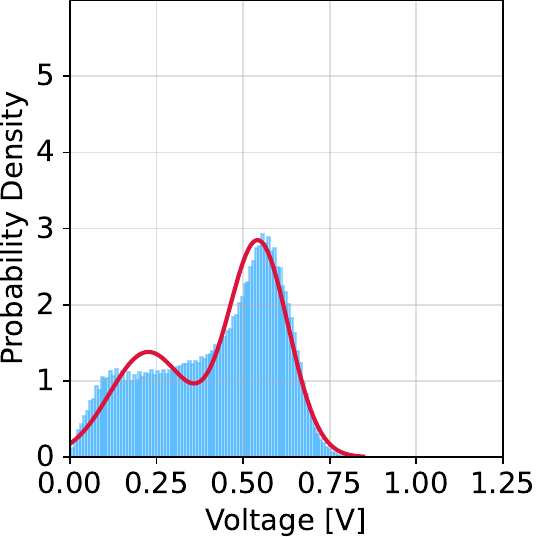}
     \caption{Blocking Card 3.}
     \label{subfig:pdf_bk3}
     \end{subfigure} 
    \hfill
    \begin{subfigure}{0.45\columnwidth}
    \centering
    \includegraphics[width=\columnwidth]{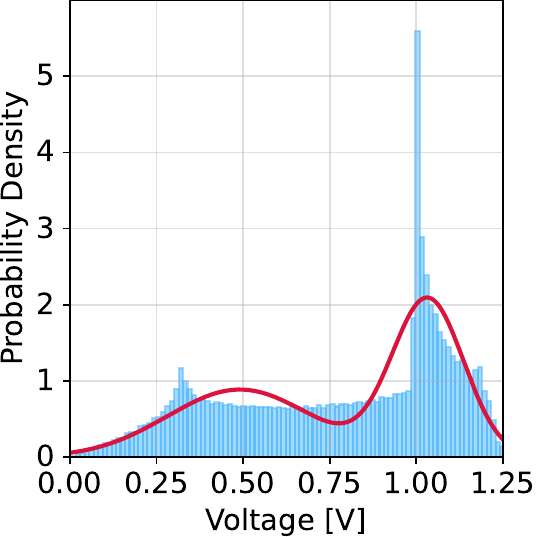}
    \caption{Blocking Card 10.}
    \label{subfig:pdf_bk10}
    \end{subfigure}
     \caption{\ac{pdf} of (a) a blocking card emitting white noise and (b) a card emitting noise at multiple frequencies.}
     \label{fig:pdf}
\end{figure}

In addition to classifying the blocking cards into passive and active ones, we further inspect their signal to identify the statistical property of the generated noise. We record the blocking card signal for about $10$ seconds to do this. We find that $9$ out of the $11$ reactive blocking cards produce nearly white Gaussian noise (see Figure~\ref{subfig:spectrum_bk3} and Figure~\ref{subfig:spectrum_bk5}), while $2$ of the reactive blocking cards produce noise signals at multiple fixed frequencies (see Figure~\ref{subfig:spectrum_bk9} and Figure~\ref{subfig:spectrum_bk10}). As an example of white noise, we refer to the power spectrum depicted in Figure~\ref{subfig:spectrum_bk3}, where the power spectral density is approximately constant, except for some peaks at fixed frequencies that include the reader's power.
All the other reactive blocking cards that generate nearly white Gaussian Noise exhibit a comparable power spectrum, as depicted in Figure~\ref{subfig:spectrum_bk5}. Although the power level may vary, the distinct peaks corresponding to the reader's power remain observable. Furthermore, Figure~\ref{subfig:pdf_bk3} confirms that the distribution can be modeled as a Gaussian mixture. 
Blocking cards that emit signals at multiple fixed frequencies, as shown in Figure~\ref{subfig:pdf_bk10}, might have a power spectrum multiplier that creates harmonics as the output of its input frequency. 
The disparity between Figure~\ref{subfig:spectrum_bk9} and Figure~\ref{subfig:spectrum_bk10} reveals that despite both cards employ a similar strategy, they utilize distinct levels of fixed frequencies.
Due to space constraints, we reported the power spectrum and \ac{pdf} generated by the other blocking cards in the code repository.

\subsection{Classification Based on Physical Inspection}
\label{subsec:shielding}
Through the spectrum analysis, we find three blocking cards that do not generate any visible noise. Furthermore, if we place these blocking cards in front of a smart card and send a message from the reader, we do not receive any reply from the smart card.
We further analyze them by inspecting their physical properties. As shown in Figure~\ref{fig:BlockingCardFlexibility} (Appendix~\ref{appendix:physics}), one of the shielding cards is very flexible, suggesting the absence of a built-in \ac{rfid} chip to emit jamming signals. After disassembling the card, as shown in Figure~\ref{fig:BlockingCardSplit} (Appendix~\ref{appendix:physics}), we notice that it has no circuit but a black layer made up of a metallic film that can block the \ac{em} field. As a result, this corroborates the idea that these cards leverage shielding materials and the Faraday principle to protect smart cards. 
Due to the specific characteristics of these cards, we faced limitations in conducting comprehensive testing using the hardware available to us. Consequently, we have made the decision to exclude these cards from the Attack Evaluation that will be conducted in Section~\ref{sec:results}.

\section{System and Threat Model}\label{sec:sysmodel}

\parag{System Model}
In our system model, we assume that the victim uses a smart card for daily activities, such as paying for the public transport used to commute to the workplace or accessing the workplace building. Being aware of the attacks that can be performed against smart cards, the user keeps a blocking card in the wallet to protect his smart card. We further assume that the victim has a smart card belonging to the ISO/IEC 14443 Type A family, which means that it can interact with \ac{nfc} readers at the $13.56$~MHz carrier frequency and a distance of up to $10cm$. In particular, the victim has either a MIFARE Ultralight or a MIFARE Classic in his wallet. 

\parag{Threat Model}
Our attacker aims at extracting the memory content of the victim's smart card even though a blocking card accompanies it. The smart card's memory content (i.e., the four or more 16 bytes blocks mentioned in Section \ref{subsec:back_classic}) depends on the specific application. Opposed to a situation where the attacker can physically steal a card or read information by looking at it, in our threat model, the attacker aims to extract information by leveraging the wireless communication capabilities of a smart card. To achieve this purpose, the attacker has to interact with the victim's smart card and follow a specific communication protocol to dump its content. 
We assume that the attacker performed some background checks on the victim so that the attacker knows which smart card is in the wallet. To save the content of the smart card, the attacker first needs an \ac{rfid} reader configured to send the set of commands compatible with the smart card communication protocol. Since the command set and series of messages for each protocol are publicly available, the attacker can easily get this information. Then, the attacker has to capture the communication between the attacker's own reader and the victim's card to access the card content upon removing the noise added by the blocking card through post-processing. For signal collection, the attacker can rely on a very cheap setup (less than $50\$$): an~\ac{rf} spectrum analyzer connected to an \ac{nfc} antenna. Alternatively, the attacker might opt for a more expensive solution, which embeds the spectrum analyzer and the antenna in a standalone device. To perform the attack, the attacker has to be physically close to the victim's wallet, although not needing a line of sight view of the victim's card. Thus, we consider that the attacker can take advantage of a scenario where the victim's wallet is in a static position (e.g., an office desk or restaurant table). After collecting signals emitted during the communication between the attacker's reader and the victim's smart card, the attacker analyzes them to exfiltrate the private information.

\section{Proposed Attack}
\label{sec:design}
\begin{figure}[t]
    \centering
    \includegraphics[width=0.9\columnwidth]{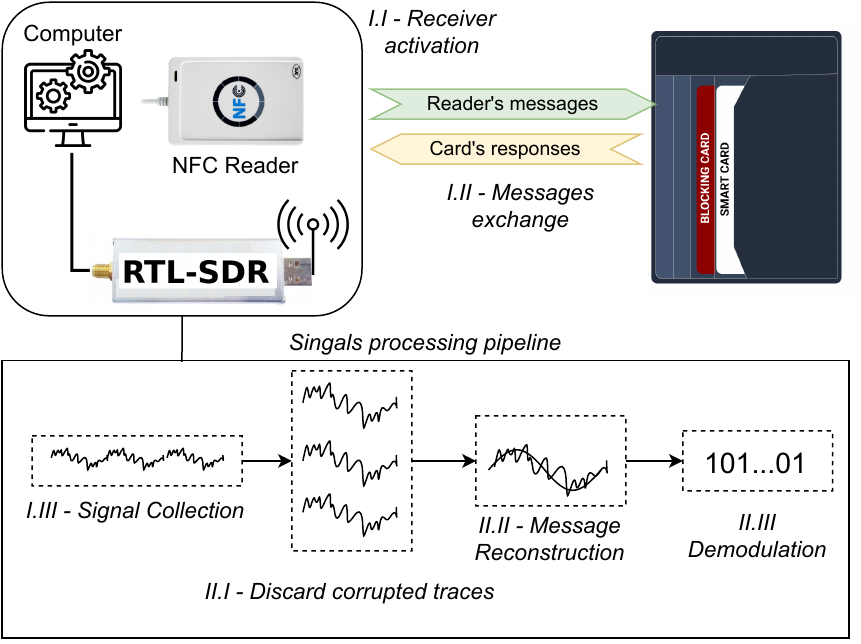}
    \caption{Setup of the attacking steps.}
    \label{fig:pipeline}
\end{figure}

Our attack aims to extract the content of the victim's smart card, even in the presence of a blocking card. Our attack can be divided into two phases, each depicted in the boxes in Figure~\ref{fig:pipeline}. The first phase aims at recording the signal emitted by the victim's smart card and then altered by the blocking card during the communication with the attacker's reader. The second phase refers to the signal processing performed to remove the noise from the blocking card and extract sensitive information from the smart card. The first phase encompasses only two steps (i.e., receiver activation and message exchange), while the second phase involves four steps (i.e., signal collection, discard of corrupted traces, message reconstruction, and demodulation). In the following, we provide more details of the steps.

\parag{Phase I - Step I: Receiver activation} First, we power the \ac{rfid} receiver to record the card responses. 

\parag{Phase I - Step II: Messages exchange} The NFC reader, configured by the attacker to communicate with the specific victim's smart card, starts sending a sequence of messages to the smart card to retrieve its content. 

\parag{Phase I - Step III: Signal collection} During the communication between the attacker's NFC reader and the victim's smart card, the antenna receives the signal generated by the latter. This step is repeated multiple times to obtain more signal samples to increase the attack success rate.

\parag{Phase II - Step I: Discard corrupted traces} Once the signal samples have been collected, the attacker can start processing them. In particular, the attacker first discards all the traces where the noise the blocking card introduces might generate errors during the upcoming demodulation procedure.

\parag{Phase II - Step II: Message Reconstruction} The attacker elaborates the collected signals to reconstruct the original message. In particular, the attacker first extracts portions of the signal where the \ac{em} field has been activated. Then, the attacker splits the entire signal into single communication sessions and discriminates the messages belonging to the smart card from the reader's ones.

\parag{Phase II - Step III: Demodulation} After successfully reconstructing the signal, the attacker demodulates it.

\section{Attack Implementation}
\label{sec:implementation}
In this section, we first describe the instrumentation we choose for our attack (i.e., Section~\ref{ssec:instrumentation}) and the implementation details of the general attack components (i.e., Section~\ref{ssec:attack_general}). We then illustrate the customization we introduce to launch the attack against a MIFARE Ultralight (i.e., Section~\ref{ssec:attack_ultralight}) and a MIFARE Classic (i.e., Section~\ref{ssec:attack_classic}). 

\subsection{Instrumentation}
\label{ssec:instrumentation}
To implement our attack in a real-world scenario, we need two main components: the first is to activate the victim's smart card and communicate with it; the second is to record the communication signal generated by the smart card and the blocking card while the communication with the first component is ongoing. We chose an ACR122U reader based on the NXP PN532 module as our first component and an RTL-SDR with a DPL-FANT antenna~\cite{le2016rfid} as the second. The following setup implies that signal processing and demodulation are performed offline after signal acquisition. However, the setup can be easily adapted to a real-time attack.

\subsection{General Implementation}
\label{ssec:attack_general}

\parag{Message exchange}
To properly configure the ACR122U reader to communicate with the victim's smart card, we must write the PN53x chips at the bit level. Thus, we select the \texttt{libnfc}
The open-source library enables the reader to send low-level commands to the smart cards~\cite{verdult2011practical}. Since the \texttt{libnfc} library triggers a segmentation fault error when there is the presence of a blocking card, we modified the library to support blocking card collision. The reader sends the same sequence of messages $80$ times to allow multiple collections of signals. The time to record $80$ communication instances is about $10$ seconds. 

\parag{Signal collection} 
To collect the communication signal generated by the smart and blocking cards, we designed the GNURadio schema (available in the repository) to control the RTL-SDR. To obtain reliable recordings in the \ac{nfc} frequency range (i.e., between $0$ and $14.4$~MHz), we modify the RTL-SDR hardware~\cite{rtl_sdr_hwmod, rtl_sdr_internals} to enable direct sampling, bypass the tuner, and enable the recording of cleaner signals in the high-frequency range. As a drawback, after this hardware modification, the \ac{rf} gain is not adjustable, and the RTL-SDR must be at a proper distance from the reader to avoid saturation. As shown in the GNURadio schema, during capture, the signal is filtered by a low-pass filter at a frequency of $f_{\rm cutoff} = f_c \pm \frac{f_{s}}{2} = 13.56MHz \pm 423.75kHz$. After filtering, we calculate the magnitude of the signal.

\parag{Discard corrupted traces}
The collected signal is processed through a Python script to remove the noise from the blocking card and retrieve the content of the smart card. To discard the corrupted traces, we verify that the signal values fall below a threshold we identified via signal inspection. If the signal values exceed this, then we discard the trace since we cannot process and demodulate it.
We also discard a trace if the standard deviation of the signal varies with respect to the standard deviation of the noise, i.e., if $std(possible\_tag) - std(noise) > 0.01$. The rationale underlying this choice is that a slight change in the standard deviation between the possible card signal and the noise allows us to detect the presence of the card signal.

\parag{Message Reconstruction}
We detect the activation zone of the \ac{em} field by applying a moving average technique given the previously defined signal range. 
Then, we establish message synchronization based on a sliding window algorithm. 
We calculate the signal gradient in each window to detect changes in the signal and, therefore, the potential start and end of a message. Finally, we match each detected message with its sender (reader or card) using two pre-determined thresholds identified by signal inspection. 

\parag{Demodulation} After reconstructing the signal, we demodulate it. The demodulator checks if the bits carried in the signal belong to the protocol patterns defined in ISO/IEC 14443 (e.g., Manchester coding modulation).
A message sequence generally starts with a start bit, continues with message bits, and is followed by an end bit. The start and end bits of the demodulated sequence are cut, and the bits are inverted according to the ISO/IEC 14443 protocol. The demodulator implementation is available in the repository.

\subsection{Customization to Attack the MIFARE Ultralight}
\label{ssec:attack_ultralight}
The MIFARE Ultralight has no authentication or encryption method. Thus, all communication signals are identical among multiple sessions. Moreover, we assume that this has no password set so that we can read all the card's content. To set up the attack against MIFARE Ultralight, we craft the ACR122U reader to send the set of messages retrieved from the official documentation of MIFARE Ultralight EV1~\cite{mifareultralightev12014n} (for the complete set of messages, refer to Table~\ref{tab:messagesUltra}, in Appendix~\ref{appendix:messagesultra}).

\subsection{Customization to Attack the MIFARE Classic}
\label{ssec:attack_classic}
Unlike MIFARE Ultralight, MIFARE Classic comes with enhanced security mechanisms. The communication is encrypted with a per-iteration different nonce; thus, the exact sequence of bytes differs from iteration to iteration. To perform the entire authentication procedure based on the CRYPTO-1 protocol, we rely on the \texttt{crapto1} library used by \ac{mfcuk}. \ac{mfcuk} is an open source C implementation~\cite{mfcuktool} of Dark Side Attack~\cite{Courtois2009TheDS}. It uses the libraries \texttt{libnfc} and \texttt{crapto1} to exploit the weakness of MIFARE Classic CRYPTO1. Similarly to Ultralight, we configure the reader with the Classic sequence of messages, taken from the official MIFARE Classic manual~\cite{mifareclassic}. For the complete set of messages, refer to Table~\ref{tab:messagesClassic}, in Appendix~\ref{appendix:messagesClassic}). 

\section{Attack Evaluation}\label{sec:results}
To evaluate the effectiveness of our attack, we considered the $11$ reactive blocking cards and a MIFARE Ultralight and a MIFARE Classic as smart cards to be protected. We specifically select MIFARE Ultralight and MIFARE Classic smart cards due to their wide adoption in several systems and lack of internal security mechanisms. We believe the main use case scenario relies on blocking cards to protect smart cards, which would otherwise be exposed to attacks. In contrast, smart cards with embedded security mechanisms, such as MIFARE Plus and MIFARE DESFire, make them more challenging to exploit.

Here, we describe our experimental setup (i.e., Section~\ref{ssec:setup}), our evaluation criteria (i.e., Section~\ref{ssec:evaluation_criteria}), and the results obtained after demodulating the signals collected from the MIFARE Ultralight (i.e., Section~\ref{ssec:results_ultralight}) and from the MIFARE Classic (i.e., Section~\ref{ssec:results_classic}). 
\subsection{Setup} 
\label{ssec:setup}
\begin{figure}[t]
    \centering
    \includegraphics[width=0.9\columnwidth]{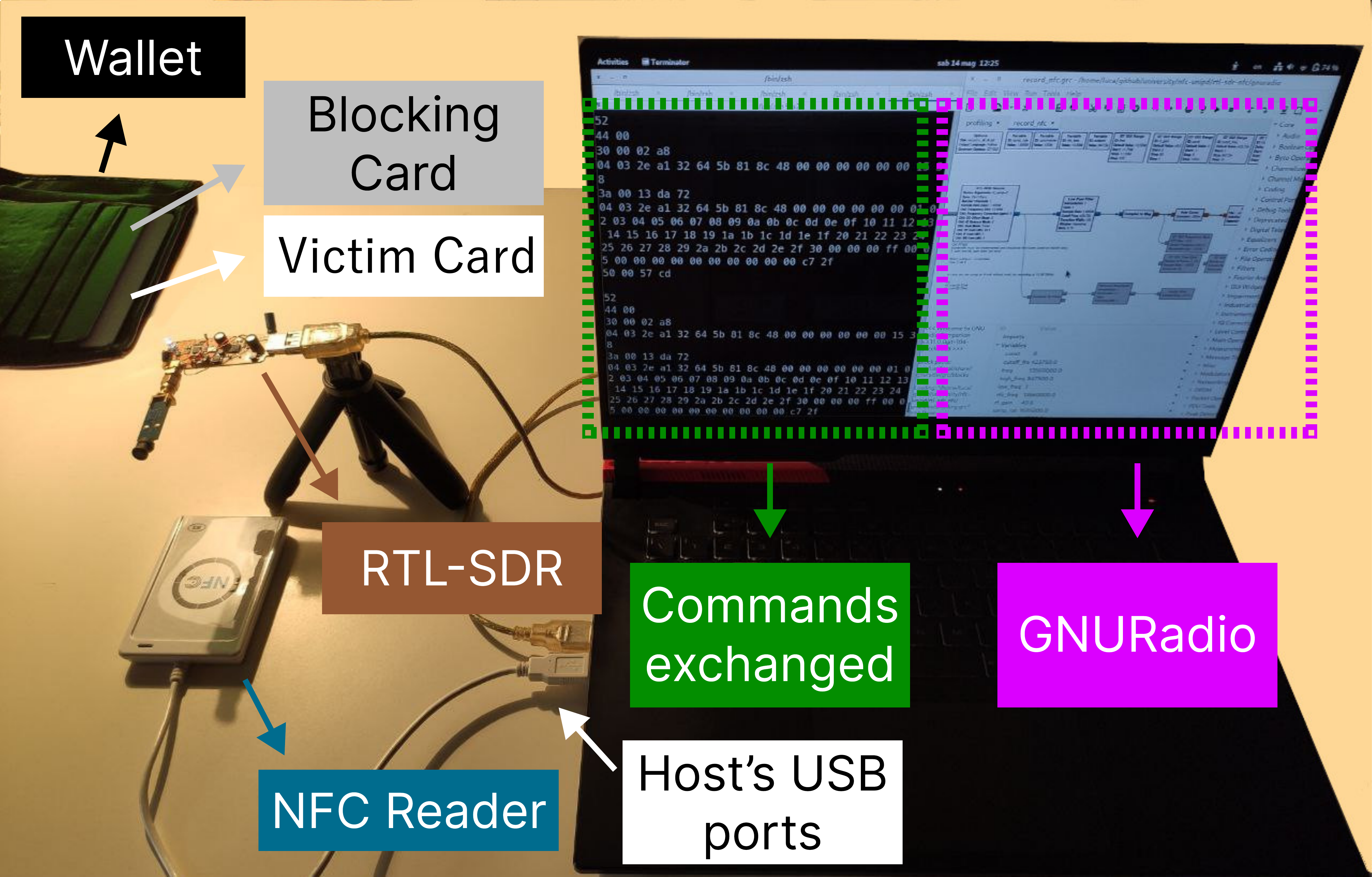}
    \caption{Setup to record the signal of an \ac{nfc} reader and card in the presence of a blocking card.}
    \label{fig:setup}
\end{figure}

Our experimental setup is shown in Figure~\ref{fig:setup}: the reader and antenna are connected to our laptop while the GNURadio program is running. Similarly to a real-world scenario\footnote{\href{https://rb.gy/z6v3i}{https://rb.gy/z6v3i}}\vphantom{d}$^{\text{,}}$\footnote{\href{https://rb.gy/tmdl8}{https://rb.gy/tmdl8}}\vphantom{d}$^{\text{,}}$\footnote{\href{https://rb.gy/4a195}{https://rb.gy/4a195}}, we place the blocking and smart card together in a wallet, in adjacent pockets. The spacing between the pockets of the wallet we use is $2.5$mm. While, according to the standard (Section~\ref{sec:smart_cards}), the reader and the smart card can communicate at a maximum distance of $10$cm, we opt for a distance of $3.5$cm between the reader and the wallet, since this is the maximum range at which the reader and the smart card in the wallet can establish a communication. In fact, the wallet generates an attenuation of the signal power.

To evaluate the effectiveness of the attacker in recovering the smart card's memory content, we write reference data into it. We then use such data as ground truth for successive steps.

\subsection{Attack Evaluation Criteria}
\label{ssec:evaluation_criteria}
We introduce three metrics to analyze the effectiveness of the $11$ blocking cards:
\begin{itemize}
    \item \textbf{Card Detection Rate}: the number of card messages detected by the attacker over the total number of card messages exchanged during the communication.
    \item \textbf{Card Demodulation Rate}: the number of card messages correctly demodulated by the attacker over the total number of card messages exchanged.
    \item \textbf{\ac{asr}}: the number of successful attack attempts over the total number of communication repetitions. The attack is considered successful if it can retrieve the content of the smart card.
\end{itemize}

These measures help us understand whether the card responds to the reader (i.e., the jamming is not effective enough to disrupt the message from the reader to the card) and whether the reader is able to correctly demodulate the response (i.e., the jamming is not effective enough to disrupt the message from the card to the reader). Therefore, thanks to these metrics, it is possible to detect if the card powers up and correctly respond.
The three above-mentioned criteria are strongly correlated with each other. In fact, an attack is complete only when the attacker detects and correctly demodulates the smart card's messages. In addition, since the attacker's goal is to read the card content, it is sufficient for the attacker to complete the reading process at least once. Therefore, we consider that a specific blocking card is successfully attacked when \ac{asr}$>0$. We summarize the results achieved with the different blocking cards in Table~\ref{tab:summary} in Appendix~\ref{appendix:summary}.

\subsubsection{MIFARE Ultralight}
\label{ssec:results_ultralight}
In Figure~\ref{subfig:card_detection_rate_ultra}, we report the Card Detection and Demodulation Rates of our attack performed against a MIFARE Ultralight, while Figure~\ref{subfig:attack_success_rate_ultra} shows the \ac{asr}. We can see that the attack is successful against $8$ blocking cards out of the $11$ analyzed, with a varying \ac{asr}. Although successful against all blocking cards that generate noise with a Gaussian mixture distribution, our attack shows a varying \ac{asr}. 

We successfully demodulate smart card messages with a percentage of $100\%$ for Blocking Cards 2 and 7. Instead, our attack does not apply to Blocking Card~6. We conjecture that this has a particular noise distribution that would be robust against our attack. Blocking Cards 9 and 10 produce signals at multiple fixed frequencies, as presented in Section~\ref{subsec:class}, and successfully prevent the demodulation of any exchanged message. Concerning Blocking Card~9, sometimes the communication starts, but one of the subsequent messages is corrupted by noise, thus disrupting the communication session. Furthermore, the RTL-SDR cannot detect any responses from the smart card. Concerning Blocking Card~10, the smart card does not respond in most cases when the reader sends a message. Due to the noise introduced by the blocking cards, the smart card cannot reply to or receive a message from the reader.

We recall that we carry out the attack by collecting $80$ communications in about $10$ seconds. The higher the number of communications collected, the more likely the attack is to be successful. On the other hand, the attacker can achieve a non-negligible chance of success with even less time.

\subsubsection{MIFARE Classic}
\label{ssec:results_classic}
\begin{figure*}[t]
    \centering
    \begin{subfigure}{0.42\textwidth}
        \centering
        \includegraphics[width=\textwidth]{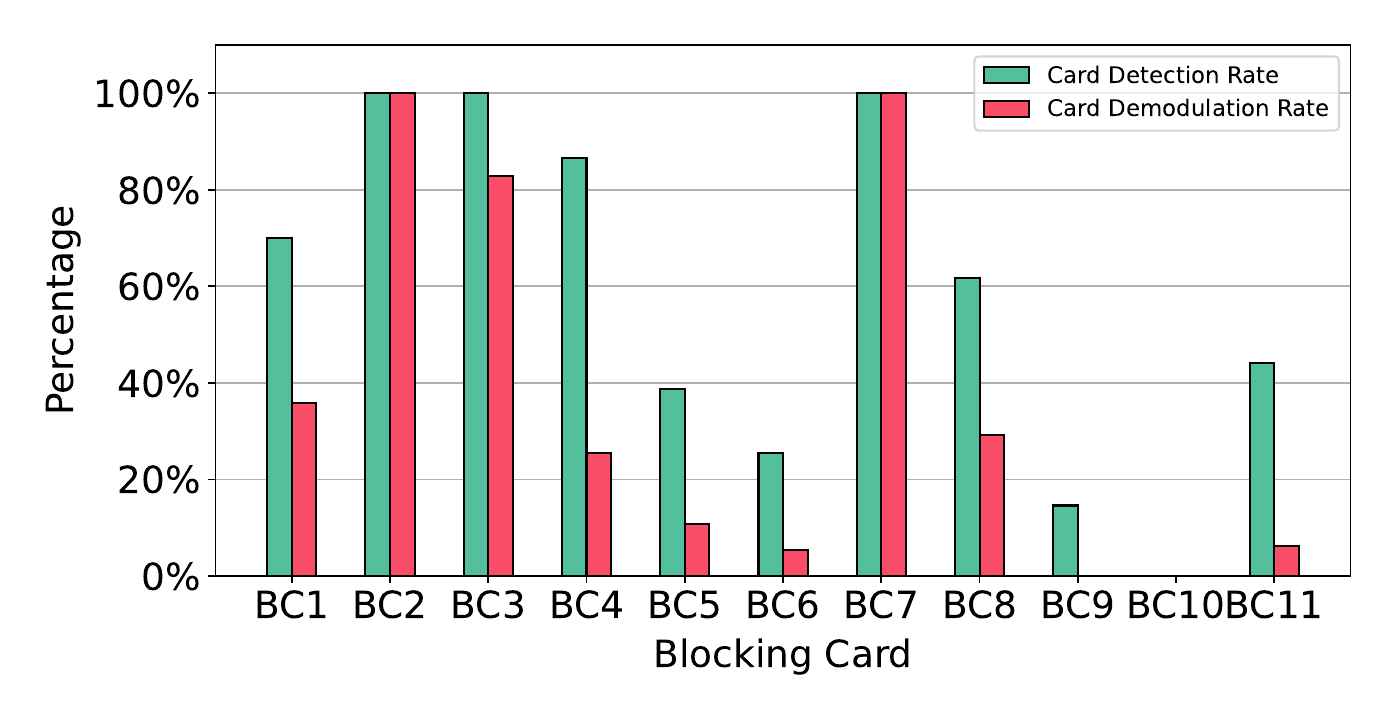}
        \caption{MIFARE Ultralight.}
        \label{subfig:card_detection_rate_ultra}
    \end{subfigure}
    \hfill
    \begin{subfigure}{0.42\textwidth}
        \centering
        \includegraphics[width=\textwidth]{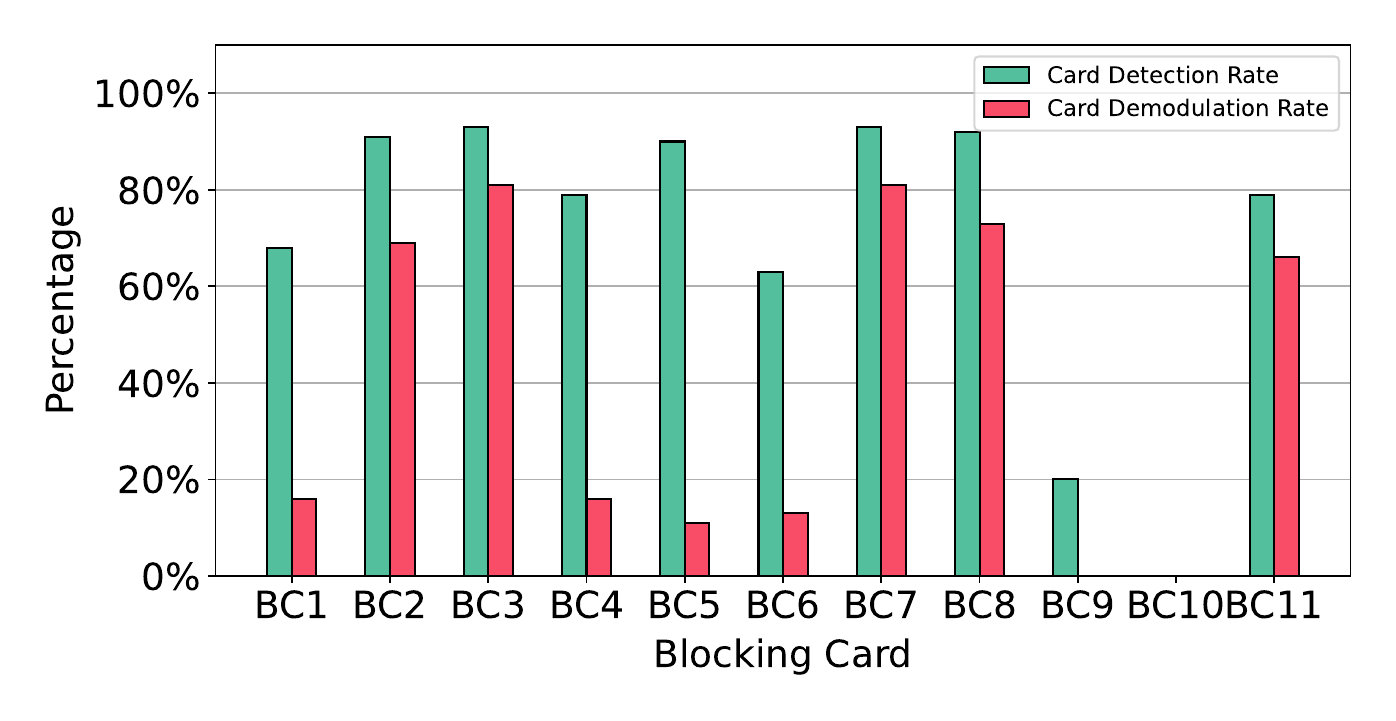}
        \caption{MIFARE Classic.}
        \label{subfig:card_detection_rate_class}
    \end{subfigure}
\caption{Card Detection and Demodulation Rates of the attacks on the different cards.}
\label{fig:results_cdr}
\end{figure*}

In Figure~\ref{subfig:card_detection_rate_class}, we report the Card Demodulation Rate and the Card Detection Rate obtained after executing the attack with the MIFARE Classic smart card. The Card Detection Rate is always above $60\%$ for most of the blocking cards following a white noise approach, which means that MIFARE Classic almost always manages to respond to the reader messages. However, the Card Demodulation Rate drops below $20\%$ with four blocking cards, which means that the demodulator cannot always demodulate the messages sent by the card. In the white noise cases, both the Card Detection Rate and Card Demodulation Rate are high, allowing us to retrieve almost always the entire communication between the card and the reader. The results of Blocking Cards 9 and 10 are similar to the MIFARE Ultralight scenario since they successfully protect the MIFARE Classic, thus not allowing the card to reply to the messages the reader sends. In fact, none of the card messages is correctly demodulated, and a small number of card messages are detected when analyzing the captured raw signal. This means that Blocking Cards 9 and 10 successfully protect the MIFARE Classic, not allowing the card to reply to the messages sent by the reader.

\begin{figure*}[t]
    \centering
    \begin{subfigure}{0.42\textwidth}
        \centering
        \includegraphics[width=\textwidth]{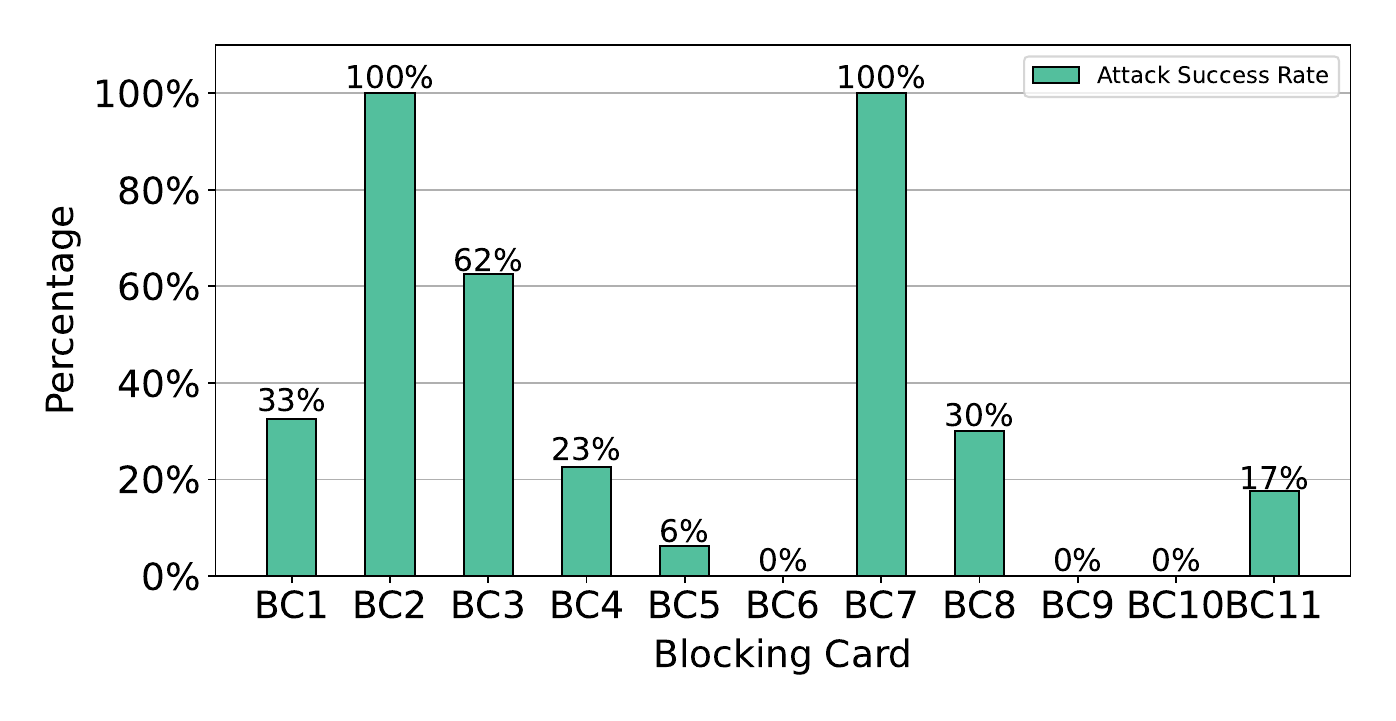}
        \caption{MIFARE Ultralight.}
        \label{subfig:attack_success_rate_ultra}
    \end{subfigure}
    \hfill
    \begin{subfigure}{0.42\textwidth}
        \centering
        \includegraphics[width=\textwidth]{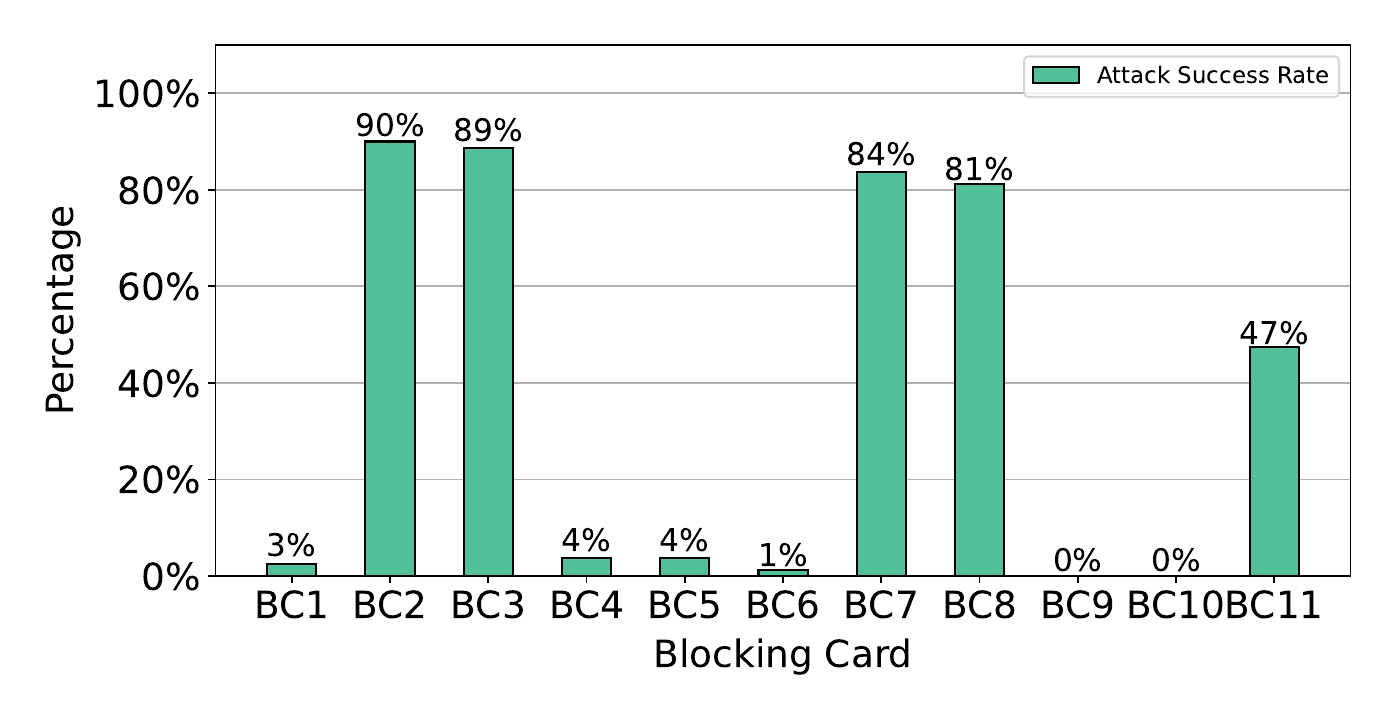}
        \caption{MIFARE Classic.}
        \label{subfig:attack_success_rate_class}
    \end{subfigure}
\caption{Attack Success Rate of the attacks on the different cards.}
\label{fig:results_asr}
\end{figure*}

Figure~\ref{subfig:attack_success_rate_class} reports the \ac{asr}.
We obtain an \ac{asr} greater than 80\% for five different blocking cards. Unlike the previous case, we can read the smart card when protected by the Blocking Card~6 (i.e., $1$ success over the $80$ communication). 

Overall, in the experiments performed with the MIFARE Classic, we obtained similar results to the scenario with the MIFARE Ultralight. We can complete the attack against the $9$ blocking cards out of the $11$ evaluated ones. However, to extract the content of a MIFARE Classic, the attacker has to send more messages than the MIFARE Ultralight, thus increasing the probability that the noise added by the blocking card corrupts a message and denies the rest of the communication. Consequently, on average, this reduces the \ac{asr}. Once again, the attacker has to find a trade-off between the time spent performing the attack and the success rate.

\subsection{Demodulation Improvement}
\label{ssec:demodulation}
\begin{figure}[t]
    \centering
    \includegraphics[width=\columnwidth]{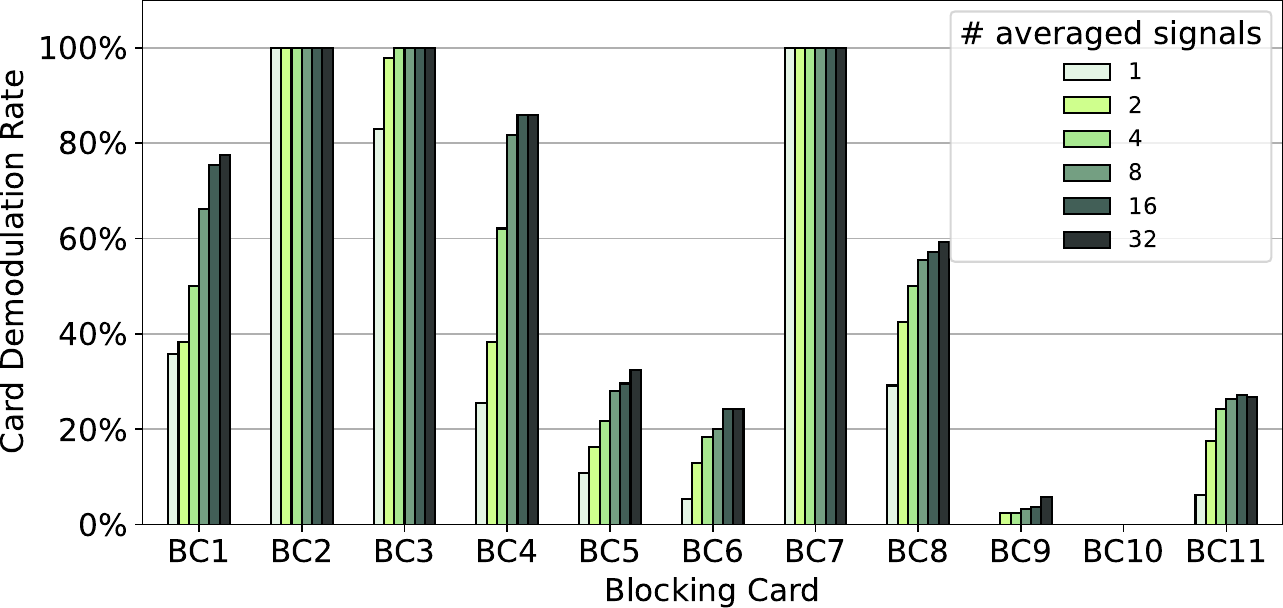}\label{subfig:blck_comparison}
    \caption{Comparison of Card Demodulation Rate on active and reactive cards by varying the signals averaged. $1$ averaged signal indicates a single signal.}
    \label{fig:evauation}
\end{figure}

To improve the attack performance, we consider reconstructing the original signal with a signal processing technique. In particular, since the noise generated by some of the blocking cards exhibits a close-to-white spectral distribution. The optimal solution for noise removal is signal averaging~\cite{hassan2010reducing}. Following this procedure, we compute the average of multiple recorded signals and try to reconstruct the original message. However, we only use this technique with the MIFARE Ultralight since it does not use any random values in the communication.
In Figure~\ref{fig:evauation}, we report the results using different numbers of averaged signals. As we can notice, by averaging multiple signals, we have a performance improvement in terms of Card Demodulation Rate in some cases. These cases correspond to a blocking card that emits white noise. On the contrary, we have no, or slight, improvement for a blocking card that emits noise at multiple frequencies (i.e., Blocking Cards 9 and 10). 
We can observe that in all blocking cards, except for Blocking Cards 9 and 10, it is possible to reconstruct the original signal with a single signal. This means that the noise generated by the blocking card, in some cases, is not sufficient to disrupt the original message.
This finding motivates the application of the attack with the MIFARE Classic. Indeed, if we can demodulate the original message with non-zero probability with only one signal, it is possible to target cards using a freshness mechanism in the authentication phase, i.e., nonces.


\section{Countermeasure}\label{sec:countermeasure}
As an outcome of our spectrum analysis performed against the $11$ reactive blocking cards, we identify two different behaviors adopted to protect a smart card: a noise emitted at multiple fixed frequencies; a noise emitted with a close-to-white spectral distribution. The first is successful, as it effectively protects smart cards from malicious attempts to steal their content. In contrast, the second approach allows an adversary to complete an attack like the one we propose in this paper. In this section, we perform an extended analysis of the features that a noise emitted by a blocking card should have to be effective, whether it is a \emph{white} or a \emph{noise emitted at multiple fixed frequencies}. To this purpose, we develop a simulated environment where the noise is directly added to the clean communication signal between the reader and the smart card without considering real-world phenomena such as attenuation or reflection. Then we consider the performance of our demodulator with different noise compositions. We experiment with clear communication recorded with a MIFARE Classic.
We measure the performance of the demodulation in terms of the Card Demodulation Rate (introduced in Section~\ref{sec:results}) and the Reader Demodulation Rate, which can be defined similarly as the number of reader messages correctly demodulated by the attacker over the total number of reader messages exchanged.

\begin{figure}[t]
\centering
\includegraphics[width=\columnwidth]{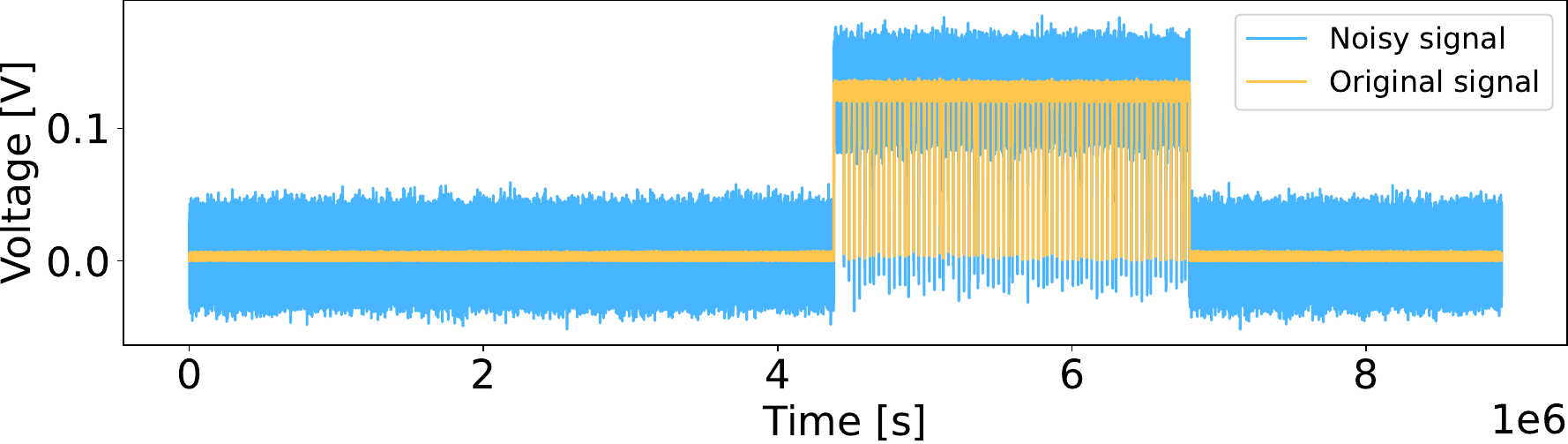}
\caption{White Gaussian noise 20\% added to a clean signal.}
\label{fig:gaussianNoise}
\end{figure}

\parag{White Gaussian noise}
In this test, we add white (i.e., uncorrelated) Gaussian noise with different compositions to the clear signal. We first calculate the \ac{std} of the original clean signal to estimate its variability. Then, we generate white Gaussian noise signal, using as \ac{std} the clean signal \ac{std} multiplied by different percentages: $5\%$, $10\%$, $15\%$, $20\%$, $25\%$, and $30\%$. Finally, we sum the original signal with the noisy signal and demodulate the message obtained with the superposition principle. Figure~\ref{fig:gaussianNoise} shows an example of the signal obtained.

\begin{figure}
\centering
\begin{subfigure}{0.48\columnwidth}
\includegraphics[width=\columnwidth]{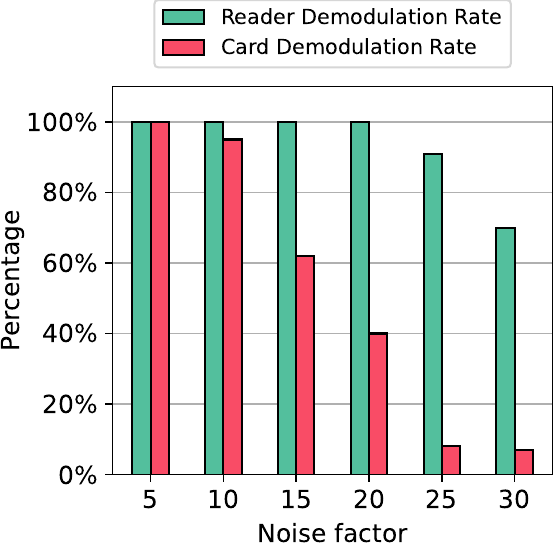}
\caption{Different Gaussian noise.}
\label{fig:gaussianDemodulationRate}
\end{subfigure}
\hfill
\begin{subfigure}{0.48\columnwidth}
\includegraphics[width=\columnwidth]{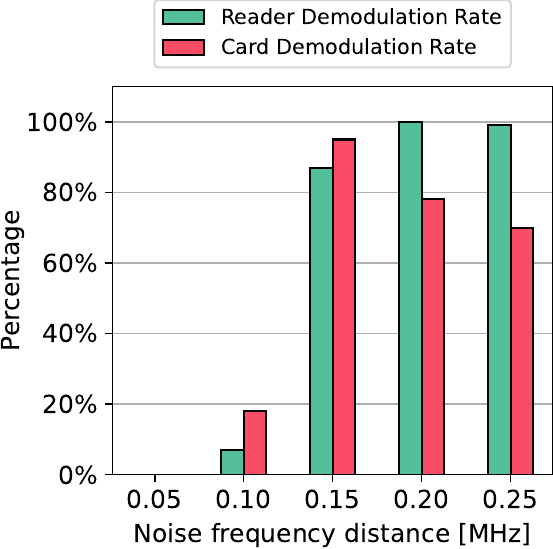}
\caption{Different peak distance.}
\label{fig:deltaDemodulationRate}
\end{subfigure}
\caption{Demodulation results with different noises.}
\label{fig:resulscounter}
\end{figure}

We report the results in Figure~\ref{fig:gaussianDemodulationRate}. The Card Demodulation Rate significantly decreases when using white Gaussian noise with a \ac{std} greater than a $15\%$ factor, whereas it dramatically drops almost to zero with a factor of $25\%$. Contrary to this, the reader Demodulation Rate is stable using a low noise factor but decreases significantly using a $30\%$ white Gaussian noise factor.

\begin{figure}
\centering
\begin{subfigure}{0.45\columnwidth}
\includegraphics[width=\columnwidth]{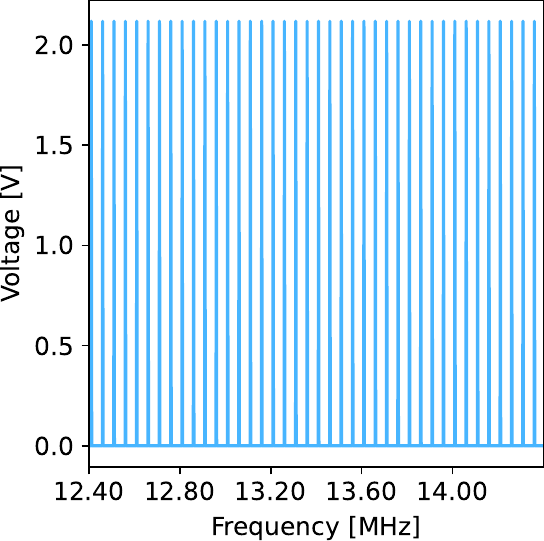}
\caption{0.05 MHz.}
\label{subfig:deltaNoise05}
\end{subfigure}
\hfill
\begin{subfigure}{0.45\columnwidth}
\includegraphics[width=\columnwidth]{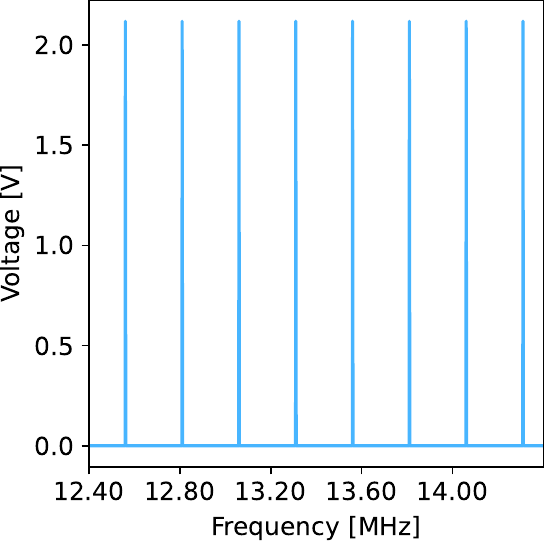}
\caption{0.25 MHz.}
\label{subfig:deltaNoise25}
\end{subfigure}
\caption{Example of noise at multiple fixed frequencies.}
\label{fig:deltaNoise}
\end{figure}

\parag{Noise at multiple fixed frequencies}
Here, we use the same setting as the previous analysis to study the effectiveness of adding noise signals at multiple fixed frequencies. More specifically, we generate different noise signals composed of peaks at equally distanced frequencies. As a frequency step (centralized at $13.56$~MHz), we experiment: $0.05$MHz, $0.10$MHz, $0.15$MHz, $0.20$MHz, $0.25$MHz.
Figure~\ref{fig:deltaNoise} reports two examples of \ac{fft} of the noise added to the clean signal.

We report the results of our demodulation performance in Figure~\ref{fig:deltaDemodulationRate}, where we can see the reader Demodulation Rate and the Card Demodulation Rate. We can observe that with a smaller interval distance, i.e., when the \ac{fft} plot is very dense, both the Reader Demodulation Rate and the Card Demodulation Rate are very low. They even reach 0\% when the distance between two consecutive peaks is only $0.05$MHz. On the contrary, when the space is above $0.150$MHz, i.e., when the \ac{fft} plot is more sparse, both the Reader Demodulation Rate and the Card Demodulation Rate are rather stable and higher than $70\%$.

\parag{Final considerations on the noise emitted by blocking cards.} Considering the results presented above, we can draw the following conclusions. In the case of \emph{white Gaussian noise}, blocking cards must add a large amount of noise to protect a smart card from malicious attacks. However, reactive blocking cards can only use the \ac{rfid} energy received from the reader to activate jamming. Therefore, there may be physical limitations in terms of the power of the signal generated. Taking into account the performance of the \emph{noise at multiple fixed frequencies}, we can claim that blocking cards should emit this type of noise with a shallow distance between two consecutive fixed frequencies. 

Finally, another strategy that blocking cards can adopt may be to randomly send bits in the transmission frequency bandwidth of the smart card: $13.56MHz\pm847.5kHz$. Similar strategies have been presented in previous works on smartphone \ac{nfc} communication~\cite{gummeson2013engarde, zhou2014nshield, di2018n}.

\section{Conclusion}\label{sec:conclusion}

In this paper, we present the first security analysis on the effectiveness of blocking cards to protect smart cards from attacks carried out over the air. We first propose a methodology to classify blocking cards according to their emitted spectrum. Our procedure allows not only to discriminate between passive and active blocking cards but also to have detailed information about the protection they adopt, which is usually information vendors do not share. 

We then select $14$ popular blocking cards on the market and perform our proposed analysis to identify their internal design. All of them are passive blocking cards, and, in particular, we find three of them adopting a shielding approach and the remaining others being reactive. To evaluate the effectiveness of such blocking cards, we designed a novel attack aimed at stealing the content of a smart card, even in the presence of a blocking card. Our proposed attack follows a methodology that can be applied to different smart cards operating with the smart card-specific communication protocol at all times. In fact, we select the MIFARE Ultralight and MIFARE Classic smart cards as the target of our attack, and we successfully manage to set it up for both of them without additional adaptions to the communication protocol.
We managed to complete our attack with $8$ blocking cards out of the $11$ reactive ones, revealing the limitations of the design of such blocking cards.
Indeed, we found that the type of noise emitted by the blocking card affects the success of the protection mechanism. The most effective noise strategy to prevent eavesdropping attempts is emitting noise at multiple fixed frequencies. However, due to the lack of details disclosed by vendors, it is difficult for a customer to select an effective blocking card.
Finally, we provide an analysis of the statistical properties of emitted noise that are effective in protecting smart cards, which may serve as a design guideline for further enhancement of blocking cards. Since no previous studies have addressed this topic, we hope that our paper will shed some light on this issue and help improve the security of blocking cards. On the other side, we plan to conduct further experiments to evaluate a higher number of blocking cards under different experimental setups (e.g., different distances between the smart card and the blocking card).


\bibliographystyle{ACM-Reference-Format}
\bibliography{biblio}

\newpage
\section*{Appendix}
\appendix

\section{Shielding Card Dissection}\label{appendix:physics}

\begin{figure}[H]
\centering
\begin{subfigure}{0.85\columnwidth}
  \centering
  \rotatebox{90}{\includegraphics[width=0.5\columnwidth]{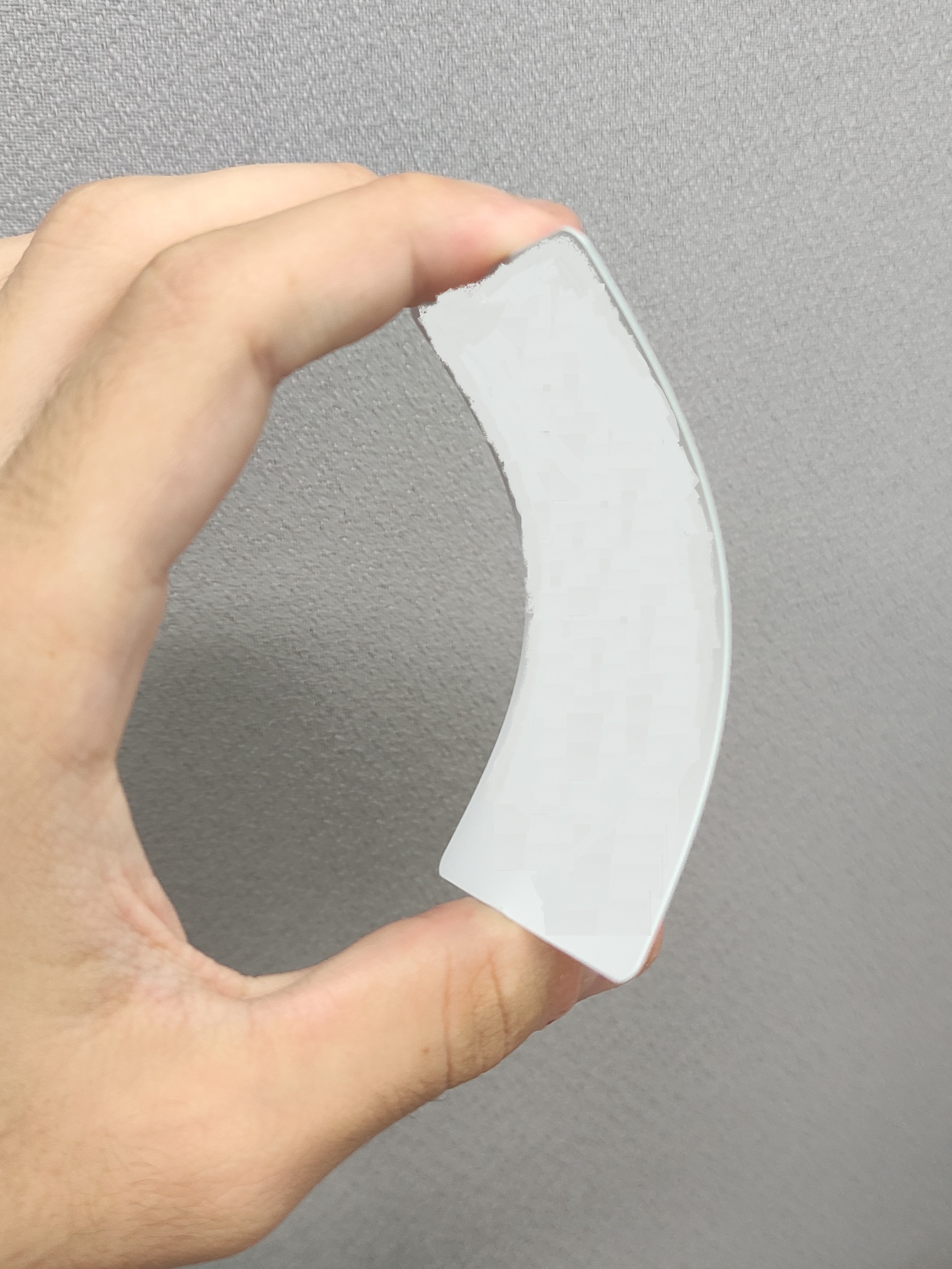}}
  \caption{Shielding card flexibility.}
    \label{fig:BlockingCardFlexibility}
\end{subfigure}

\hspace{30pt}

\begin{subfigure}{0.85\columnwidth}
  \centering
  \rotatebox{90}{\includegraphics[width=0.5\columnwidth]{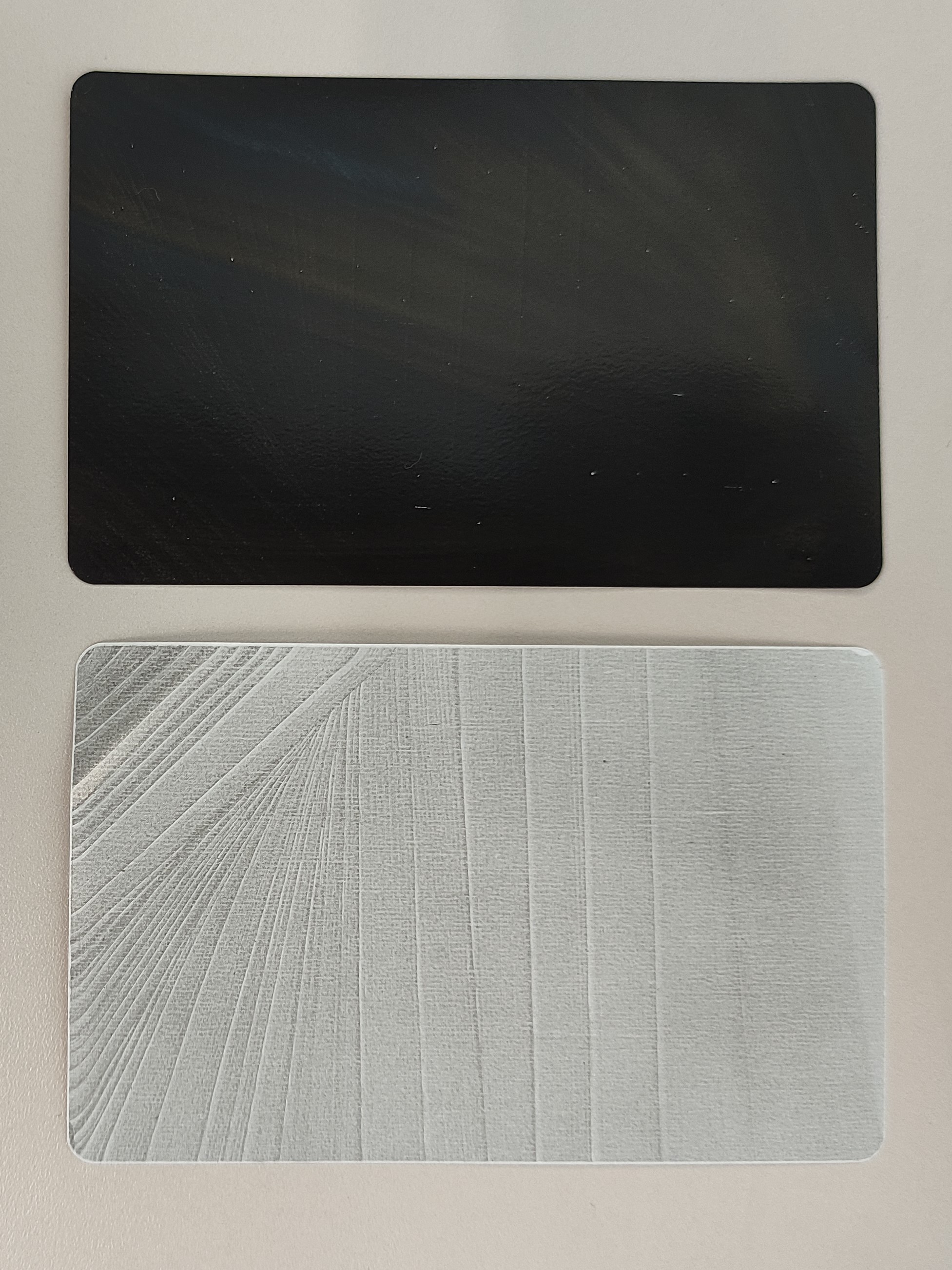}}
  \caption{Shielding card internal.}
\label{fig:BlockingCardSplit}
\end{subfigure}
\caption{Physical characteristics of a shielding card.}
\label{fig:BlockingCard}
\end{figure}

In Figure~\ref{fig:BlockingCard} we show the physical composition of a shielding blocking card that highlights the absence of an integrated circuit.

\begin{table*}[!b]
\centering
\caption{Summary of the results on the various Blocking Cards.}
\label{tab:summary}
\resizebox{\textwidth}{!}{
\begin{tabular}{|c|l|ccc|ccc|cccccc|}
\hline
\multicolumn{1}{|l|}{\textbf{}}                                      & \multicolumn{1}{c|}{\textbf{}}     & \multicolumn{3}{c|}{\textbf{MIFARE Ultralight}} & \multicolumn{3}{c|}{\textbf{MIFARE Classic}}                                                                                                                              & \multicolumn{6}{c|}{\textbf{\begin{tabular}[c]{@{}c@{}}MIFARE Ultralight\\ Demodulation Improvment\end{tabular}}}                                                                                                           \\ \cline{3-14} 
\textbf{\begin{tabular}[c]{@{}c@{}}Blocking \\ Card ID\end{tabular}} & \multicolumn{1}{c|}{\textbf{Type}} & \multicolumn{1}{c|}{\begin{tabular}[c]{@{}c@{}}Detection\\ Rate\end{tabular}} & \multicolumn{1}{c|}{\begin{tabular}[c]{@{}c@{}}Demodulation\\ Rate\end{tabular}} & ASR  & \multicolumn{1}{c|}{\begin{tabular}[c]{@{}c@{}}Detection \\ Rate\end{tabular}} & \multicolumn{1}{c|}{\begin{tabular}[c]{@{}c@{}}Demodulation \\ Rate\end{tabular}} & ASR  & \multicolumn{1}{c|}{\textit{1}} & \multicolumn{1}{c|}{\textit{2}} & \multicolumn{1}{c|}{\textit{4}} & \multicolumn{1}{c|}{\textit{8}} & \multicolumn{1}{c|}{\textit{16}} & \multicolumn{1}{c|}{\textit{32}}                 \\ \hline
1                                                                    & Gaussian                           & 0.70                                                                          & 0.36                                                                             & 0.33 & 0.68                                                                           & 0.16                                                                              & 0.03 & 0.38     & 0.38     & 0.50     & 0.66    &  0.75      & 0.78 \\ \hline
2                                                                    & Gaussian                           & 1.00                                                                          & 1.00                                                                             & 1.00 & 0.91                                                                           & 0.69                                                                              & 0.90 & 1.00                            & 1.00                            & 1.00                            & 1.00                            & 1.00                             & 1.00                        \\ \hline
3                                                                    & Gaussian                           & 1.00                                                                          & 0.83                                                                             & 0.63 & 0.93                                                                           & 0.81                                                                              & 0.89 & 0.83                            & 0.98                            & 1.00                            & 1.00                            & 1.00                             & 1.00                        \\ \hline
4                                                                    & Gaussian                           & 0.87                                                                          & 0.25                                                                             & 0.23 & 0.79                                                                           & 0.16                                                                              & 0.04 & 0.25                            & 0.38                            & 0.62                            & 0.82                            & 0.86                             & 0.86                       \\ \hline
5                                                                    & Gaussian                           & 0.39                                                                          & 0.11                                                                             & 0.06 & 0.90                                                                           & 0.11                                                                              & 0.04 & 0.11                            & 0.16                            & 0.22                            & 0.28                            & 0.30                             & 0.33                       \\ \hline
6                                                                    & Gaussian                           & 0.25                                                                          & 0.05                                                                             & 0.00 & 0.63                                                                           & 0.13                                                                              & 0.01 & 0.05                            & 0.13                            & 0.18                            & 0.20                            & 0.24                             & 0.24                       \\ \hline
7                                                                    & Gaussian                           & 1.00                                                                          & 1.00                                                                             & 1.00 & 0.93                                                                           & 0.81                                                                              & 0.84 & 1.00                            & 1.00                            & 1.00                            & 1.00                            & 1.00                             & 1.00                        \\ \hline
8                                                                    & Gaussian                           & 0.62                                                                          & 0.29                                                                             & 0.30 & 0.92                                                                           & 0.73                                                                              & 0.81 & 0.29                            & 0.43                            & 0.50                            & 0.55                            & 0.57                             & 0.59                        \\ \hline
9                                                                    & Fixed frequencies                  & 0.01                                                                          & 0.00                                                                             & 0.00 & 0.20                                                                           & 0.00                                                                              & 0.00 & 0.00                            & 0.03                            & 0.03                            & 0.03                            & 0.04                             & 0.06                                             \\ \hline
10                                                                   & Fixed frequencies                  & 0.00                                                                          & 0.00                                                                             & 0.00 & 0.00                                                                           & 0.00                                                                              & 0.00 & 0.00                            & 0.00                            & 0.00                            & 0.00                            & 0.00                             & 0.00                        \\ \hline
11                                                                   & Gaussian                           & 0.44                                                                          & 0.06                                                                             & 0.18 & 0.79                                                                           & 0.66                                                                              & 0.48 & 0.06                            & 0.18                            & 0.24                            & 0.26                            & 0.27                             & 0.27                        \\ \hline
12                                                                   & Shielding                          & -                                                                             & -                                                                                & -    & -                                                                              & -                                                                                 & -    & -                               & -                               & -                               & -                               & -                                & -                           \\ \hline
13                                                                   & Shielding                          & -                                                                             & -                                                                                & -    & -                                                                              & -                                                                                 & -    & -                               & -                               & -                               & -                               & -                                & -                           \\ \hline
14                                                                   & Shielding                          & -                                                                             & -                                                                                & -    & -                                                                              & -                                                                                 & -    & -                               & -                               & -                               & -                               & -                                & -                           \\ \hline
\end{tabular}
}
\end{table*}

\section{MIFARE Ultralight Message Set}\label{appendix:messagesultra}
Table~\ref{tab:messagesUltra} reports the list of messages exchanged between the reader and the MIFARE Ultralight during the signal acquisition~\cite{mifareultralightev12014n}.

\begin{table}[H]
\caption{Messages that were repeatedly exchanged (80 times) between the reader (R) and the card (C).}
\label{tab:messagesUltra}
\centering
\resizebox{0.8\columnwidth}{!}{
\begin{tabular}{|l|l|l|l|}
\hline
Device   &   Command           & Arg1           & Arg2                      \\ \hline
R  & WUPA & & \\ \hline
C & ATQA & & \\ \hline
R & READ from page 00h & 00 (Addr) & \\ \hline
C  & result of READ & & \\ \hline
R & FAST READ & 00h (StartAddr) & 13 (EndAddr) \\ \hline
C & FAST READ result & & \\ \hline
R & HALT & & \\
\hline
\end{tabular}
}
\end{table}

\section{MIFARE Classic Message Set}\label{appendix:messagesClassic}
Table~\ref{tab:messagesClassic} reports the list of messages exchanged between the reader and the MIFARE Classic during the signal acquisition. 

\begin{table}[H]
\centering
\caption{Set of messages exchanged between the reader (R) and the MIFARE Classic (C).}
\label{tab:messagesClassic}
\resizebox{0.5\columnwidth}{!}{
\begin{tabular}{|l|l|l|l|l|}
\hline
\multicolumn{1}{|c|}{\textbf{Step}} &
  \multicolumn{1}{c|}{\textbf{Device}} &
  \multicolumn{1}{c|}{\textbf{Command}} \\ \hline
1  & R & WUPA                                       \\ \hline
2  & C & REQA                                       \\ \hline
3  & R & SELECT                                     \\ \hline
4  & C & UID + BCC                                  \\ \hline
5  & R & SELECT + UID                               \\ \hline
6  & C & MIFARE 1K                                  \\ \hline
7  & R & AUTH (Block 0x07)                          \\ \hline
8  & C & $n_T$                                      \\ \hline
9  & R & $n_R \oplus ks_1 + a_R \oplus ks_2$        \\ \hline
10 & C & $a_T \oplus ks_3$                          \\ \hline
11 & R & READ                                       \\ \hline
12 & C & Result of READ                             \\ \hline
13 & R & HALT                                       \\ \hline
\end{tabular}
}
\end{table}

\section{Summary of Results}\label{appendix:summary}

In Table~\ref{tab:summary} we summarize the results achieved with the different blocking cards.

\end{document}